\documentclass[10pt,aps,pra,amsmath,amssymb,amsfonts,floatfix,twocolumn,superscriptaddress]{revtex4-2}
\usepackage[T1]{fontenc}
\usepackage[utf8]{inputenc}
\usepackage{graphicx}
\usepackage{bm}
\usepackage[separate-uncertainty]{siunitx}
\usepackage[breaklinks=true]{hyperref}
\hypersetup{
colorlinks=true,
linkcolor=blue,
citecolor=blue,
urlcolor=blue
}

\begin{document}

\title{Brilliant multi-GeV Compton \texorpdfstring{$\gamma$}{gamma}-ray source seeded by a photon accelerator}

\author{Michael~J.~Quin}
\email{michael.quin@physics.gu.se}
\affiliation{Department of Physics, University of Gothenburg, SE-41296 Gothenburg, Sweden}

\author{Stepan~S.~Bulanov}
\affiliation{Lawrence Berkeley National Laboratory, Berkeley, California 94720, USA}

\author{Arkady~Gonoskov}
\affiliation{Department of Physics, University of Gothenburg, SE-41296 Gothenburg, Sweden}

\author{Christopher~D.~Murphy}
\affiliation{York Plasma Institute, School of Physics, Engineering and Technology, University of York, York, UK}

\author{Mattias~Marklund}
\affiliation{Department of Physics and Astronomy, Chalmers University of Technology, SE-41296 Gothenburg, Sweden}
\affiliation{Department of Physics, University of Gothenburg, SE-41296 Gothenburg, Sweden}

\author{Alexander~G.~R.~Thomas}
\affiliation{G{\'e}rard Mourou Center for Ultrafast Optical Sciences, University of Michigan, Ann Arbor, Michigan 48109, USA}

\author{Thomas~G.~Blackburn}
\email{tom.blackburn@physics.gu.se}
\affiliation{Department of Physics, University of Gothenburg, SE-41296 Gothenburg, Sweden}

\date{\today}

\begin{abstract}

High-brilliance sources of polarized $\gamma$ rays are widely sought after to pump and probe matter at sub-atomic length scales.
However, existing accelerator facilities and optical lasers cannot reach a sufficiently high center-of-mass energy to produce polarized, multi-\si{\giga\electronvolt} $\gamma$ rays from unpolarized electrons via inverse Compton scattering.
Here we propose a scheme where the optical laser photons are first ``accelerated'' to the extreme ultraviolet in a beam-driven plasma wakefield, then reflected by a plasma mirror back onto a trailing electron beam, producing a flash of $\gamma$ rays.
Numerical simulations demonstrate this light source can achieve a high peak-brilliance ($10^{25}$photons/\si{\second\,\milli\metre^2\,\milli\radian^2}\,0.1\%\,BW) and a high degree of circular (95\,\%) or linear (77\,\%) polarization at multi-\si{\giga\electronvolt} photon energies, paving the way for the production of spin-polarized positrons and tests of light-by-light scattering.

\end{abstract}

\maketitle

\section{Introduction}

High-energy photons ($\gamma$ rays) can be generated by inverse Compton scattering when optical or ultraviolet photons collide with and gain energy from ultrarelativistic electrons~\cite{weller2009, yan2019}. Highly polarized multi-\si{\giga\electronvolt} $\gamma$ rays are indispensable when probing the structure of sub-atomic matter via hadron spectroscopy and searching for quantum chromodynamical exotics such as hybrid mesons and glueballs~\cite{adhikari2021, muramatsu2022}. Looking ahead, a brilliant multi-\si{\giga\electronvolt} $\gamma$-ray source would pave the way for new tests of fundamental physics, including the production of spin-polarized positrons for a future lepton collider~\cite{bulanov2023, joshi2025, abramowicz2026}, the observation of light-by-light scattering~\cite{marklund2006, king2016}, and the search for axion-like particles~\cite{bai2022, abramowicz2021}.

In this paper, we show how brilliant multi-\si{\giga\electronvolt} $\gamma$ rays can be generated at existing facilities which co-locate a linear accelerator and a few-\si{\tera\watt} optical laser~\cite{yakimenko2019, abramowicz2021}. As current accelerators are limited to electron energies close to 10\,\si{\giga\electronvolt}, kinematics requires that the incident photon energy must lie in the extreme ultraviolet (XUV) in order to produce multi-\si{\giga\electronvolt} $\gamma$ rays via linear Compton scattering. In the linear regime, each electron interacts with a single laser photon, in contrast to nonlinear Compton scattering where many laser photons can be absorbed coherently to produce a hard $\gamma$ ray~\cite{dipiazza2012, gonoskov2022, fedotov2023}.

The advantage of linear Compton scattering with an XUV laser pulse, as opposed to nonlinear Compton scattering with a high-intensity optical laser pulse~\cite{mirzaie2024, matheron2024, gerstmayr2025}, is that the former imprints a high degree of linear or circular polarization onto a narrowband, collimated $\gamma$-ray beam; in the nonlinear case, $\gamma$-ray emission is broadband, with a wide angular divergence, and restricted to linear polarization unless the electrons are initially spin-polarized~\cite{wu2025}.

Intense XUV light can be generated by reflecting an optical laser pulse from a surface plasma~\cite{chopineau2021, timmis2026} or flying mirror~\cite{bulanov2003, lamac2026_prr}.
However these mechanisms are inherently broadband, and therefore unsuitable for seeding a Compton $\gamma$-ray source. Instead, a monochromatic XUV pulse can be generated in a few millimeters by `photon acceleration' of an optical pulse as it copropagates with the moving refractive-index gradient of a plasma wakefield~\cite{wilks1989, mendonca2000, murphy2006, esarey2009}.
By employing a tailored plasma-density down ramp~\cite{sandberg2023_prl, sandberg2023_pre}, frequency upshifts of $\times22$ have been demonstrated by particle-in-cell (PIC) simulations of a plasma-wakefield photon accelerator driven by a 50-\si{\giga\electronvolt} electron beam~\cite{miller2025, russell2025}.

\begin{figure}[t]
    \includegraphics[width=\linewidth]{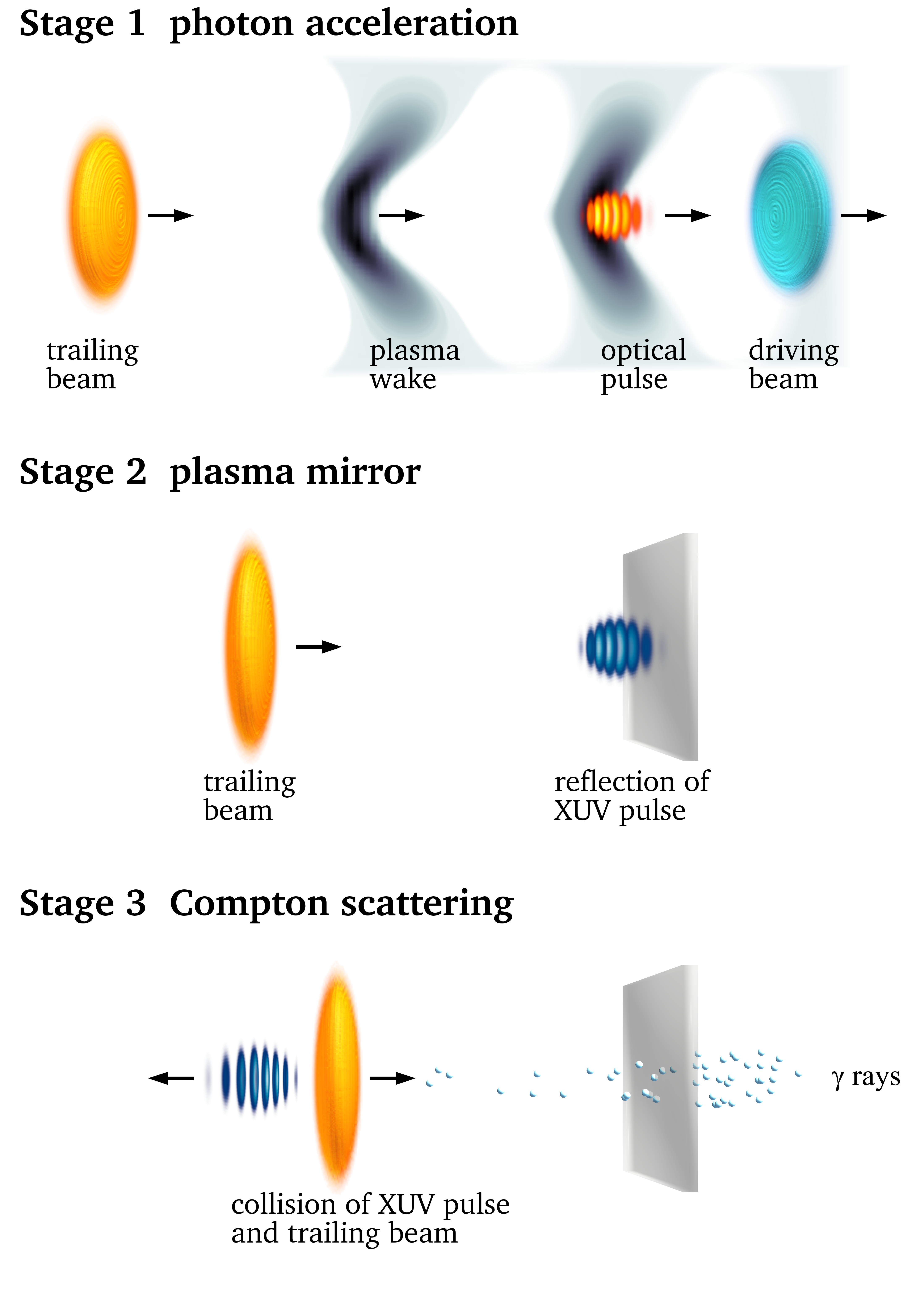}
    \caption{Schematic of photon-accelerator-seeded Compton $\gamma$-ray source. (Stage 1) Photon acceleration in an electron-beam-driven plasma wakefield. (Stage 2) Driver and XUV pulse impinge on plasma mirror, laser pulse is back-reflected while the driver propagates through undeflected. (Stage 3) Collision of reflected XUV pulse and trailing beam produces a flash of Compton $\gamma$ rays. Laser pulse and plasma wake were taken from PIC data shortly after the beginning of the simulation.
    }
    \label{fig:schematic}
\end{figure}

Here, we consider an optical laser pulse propagating between two 10-\si{\giga\electronvolt} electron bunches, one driving and one trailing, as employed for plasma-wakefield acceleration of electrons~\cite{storey2024, lindstrom2025}.
The optical pulse is accelerated to XUV frequencies in the plasma-wakefield induced by the driver, and then back-reflected~\cite{taphuoc2012, tsai2015} by a plasma mirror onto the trailing beam, as shown in Fig.~\ref{fig:schematic}.
By modeling the collision of the XUV pulse and trailing beam in the locally monochromatic approximation of quantum electrodynamics (QED)~\cite{heinzl2020, blackburn2023}, we demonstrate that a Compton $\gamma$-ray source seeded by a photon accelerator can achieve a high peak brilliance ($10^{25}$photons/\si{\second\,\milli\metre^2\,\milli\radian^2}\,0.1\%\,BW) and a high degree of circular (95\,\%) or linear (77\,\%) polarization at the peak (6.7\,\si{\giga\electronvolt}) before the Compton edge, limited only by recoil from photon emission.
The location of the Compton edge, defined as the maximum photon energy permitted by kinematics, can be varied by changing the plasma propagation distance, while the $\gamma$-ray polarization is efficiently inherited from the initial laser pulse.

\begin{figure*}[t]
    \includegraphics[width=\linewidth]{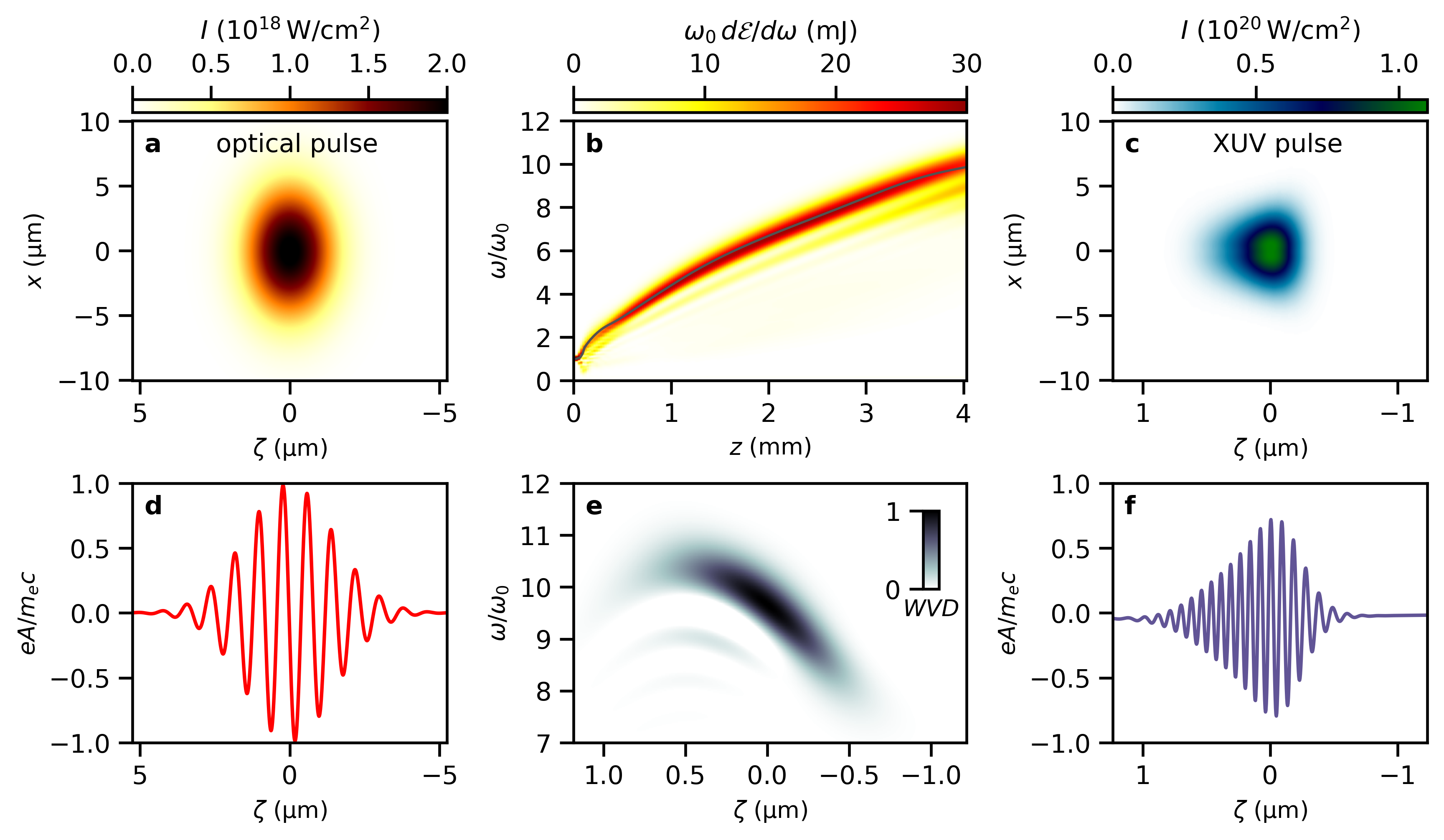}
    \caption{Evolution of linearly polarized laser pulse during photon acceleration. (a) Transverse profile of initial laser pulse. (b) Energy-spectrum and central frequency (grey line) while propagating in the wake. (c) Transverse profile of final laser pulse. (d) Vector potential of initial laser pulse. (e) Wigner-Ville distribution and (f) vector potential of final laser pulse, showing a significant down-chirp and asymmetric temporal profile. $\zeta=ct-z$ is the light-front coordinate. Results are identical for circular polarization, except the initial and final amplitude are decreased by $\sqrt{2}$. The final XUV laser pulse is relativistically intense, with a normalized amplitude in the order of unity.
    }
    \label{fig:photon_acc}
\end{figure*}

\section{Photon acceleration}

PIC simulations are performed to demonstrate that a $\times10$ frequency upshift can be achieved by a plasma-wakefield photon accelerator driven by a 10\,\si{\giga\electronvolt} electron beam.
The driver contains charge $Q=4.7\,\si{\nano\coulomb}$, has a flat longitudinal profile $l=0.3\,\si{\micro\meter}$ and a wide Gaussian transverse profile $\sigma_r=14\,\si{\micro\meter}$ in order to efficiently accelerate the laser pulse. A linearly-polarized (circularly-polarized) 14-\si{\milli\joule} laser pulse is located in the first wavefront behind the driver with normalized amplitude $a_0=1$ ($a_0=1/\sqrt{2}$), central wavelength $\lambda_0=800\,\si{\nano\meter}$, and FWHM duration $\tau_0=8.2\,\si{\femto\second}$. Here the normalized amplitude $a_0=eE_0/m_e\omega_0c$ is defined in terms of the central frequency $\omega_0$ and electric field amplitude $E_0$.

While the driver is ultrarelativistic, with Lorentz factor $\gamma_0\gg 1$, the laser pulse propagates at a group velocity determined by its instantaneous frequency. A tailored plasma density down ramp ensures that the laser remains in the accelerating region at all times by matching the wake velocity to the group velocity~(see Appendix~\ref{ap:photon_acc_sims}). Similar \si{\milli\meter}-scale down ramps have been realized by varying the size of an aperture between two compartments of a gas cell~\cite{kononenko2016}.

The central frequency of the laser pulse reaches the XUV domain after propagating through $z\approx4\,\si{\milli\meter}$ of plasma (Fig.~\ref{fig:photon_acc}\,b). This produces a commensurate increase of the laser intensity and energy (Fig.~\ref{fig:photon_acc}\,a,\,c), indicating that a significant fraction of photons are captured and accelerated by the wake, where the number of photons is approximately conserved in an underdense plasma~\cite{mendonca2000}. The driver is short $k_{p0}l < 1$ and wide $k_{p0}\sigma_r\gg 1$ compared to the plasma skin depth, with a peak density $n_d$ equal to that of the plasma at the top of the down ramp $n_{p0}$, where we define the quiescent plasma frequency $\omega_{p0}=\sqrt{n_{p0} e^2/\varepsilon_0 m_e}$ and wavenumber $k_{p0}=\omega_{p0}/c$. In this case, the derivative of the photon frequency is proportional to the normalized areal-charge density $A=k_{p0} l n_d/n_{p0}\approx 0.5$ and local plasma density $n_p\equiv n_p(z)$~\cite{sandberg2023_pre}. As the local plasma density decreases according to the down ramp, larger frequency shifts require an increasingly long propagation distance.

The XUV pulse emitted by the photon accelerator is relativistically intense with a normalized amplitude just below unity (Fig.~\ref{fig:photon_acc}\,d,\,f). In practice only a fraction of the initial laser pulse, significantly smaller than the plasma wavelength, is captured and accelerated by the wake. The gradient of the refractive index is steeper behind the pulse, leading to a significant down-chirp, while the curvature of the wake causes the temporal profile to become asymmetric (Fig.~\ref{fig:photon_acc}\,e).

At the end of the photon accelerator, the electron driver and XUV pulse strike a plasma mirror (e.g. Kapton tape). 
The transverse electric field of the driver ionizes the surface several femtoseconds before the arrival of the XUV pulse (see Appendix~\ref{ap:plasma_mirror}). After reflection, the XUV pulse diverges while propagating back through the plasma before colliding with the trailing beam in vacuum.
For linear (circular) polarization, the XUV pulse has a normalized amplitude $a_0'\approx 0.11$ ($a_0'\approx 0.08$) and instantaneous waist $w_0'\approx 30\,\si{\micro\meter}$ at the point of collision.

\section{Collision of XUV pulse and trailing beam}

A head-on laser-electron collision is governed by two parameters in quantum electrodynamics (QED), the normalized amplitude $a_0'$ and linear quantum parameter $\eta_0'=2\gamma_0\hbar\omega_0'/m_ec^2$ for an ultrarelativistic electron $\gamma_0\gg 1$~\cite{dipiazza2012, gonoskov2022, fedotov2023}. The reflected XUV pulse is within the linear regime $a_0' < 1$, yet the high frequency $\hbar \omega_0'\approx  15.5\,\si{\electronvolt}$ indicates that the scattering is fully quantum mechanical: $\eta_0'\approx 1.2$. Consequently, nonlinear corrections to the Klein-Nishina theory of Compton scattering play a minor role, but recoil from photon emission will strongly affect the electron dynamics~\cite{landaulifshitz_vol2, ritus1985}.

\begin{figure*}[t]
    \includegraphics[width=\linewidth]{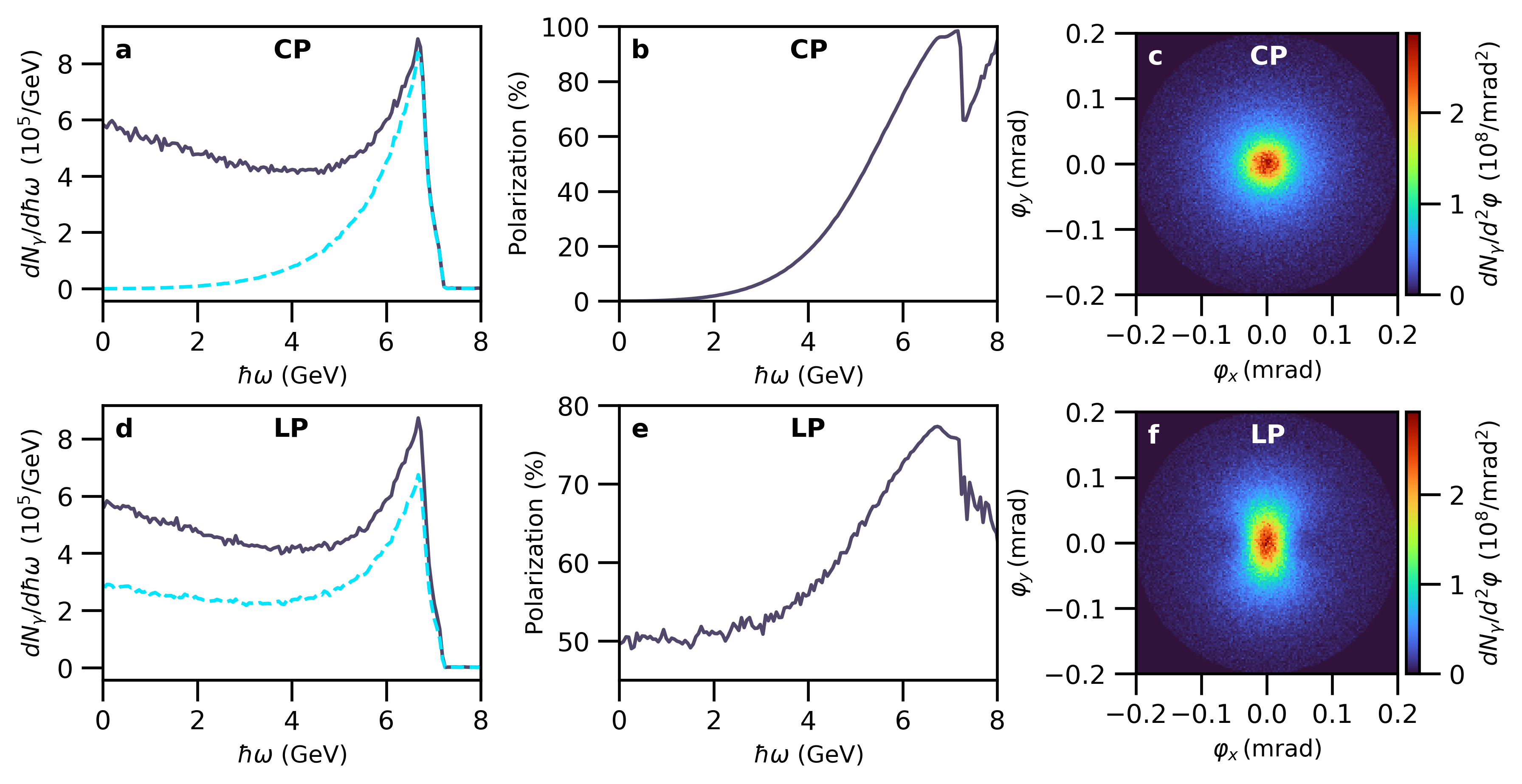}
    \caption{Compton $\gamma$ rays emitted from collision of XUV pulse and trailing beam.
    (a) Photon spectrum, (b) degree of right-handed circular polarization, and (c) photon angular distribution for collision with circularly-polarized (CP) laser pulse.
    (d) Photon spectrum, (e) degree of polarization parallel to the electric field, and (c) photon angular distribution, for collision with linearly-polarized (LP) laser pulse. 
    The solid dark-blue line indicates the properties of the $\gamma$-ray beam regardless of polarization, while the dashed light-blue line is weighted by the degree of polarization.
    }
    \label{fig:gamma-rays}
\end{figure*}

Numerical simulations are performed with the particle-tracking code Ptarmigan~\cite{blackburn2023} to model the collision of the reflected XUV pulse and trailing beam (see Appendix~\ref{ap:collision_sims}). Here, space charge effects are negligible, and the trajectories evolve according to the external field (reflected XUV pulse). The probability of a quantum mechanical event, such as Compton scattering or electron-positron pair creation, is determined by Monte-Carlo sampling of emission rates derived in the locally monochromatic approximation of QED~\cite{heinzl2020}, which are required in the regime $a_0'\sim 1$.

The spectrum of $\gamma$ rays emitted from the collision peaks just below the Compton edge $\hbar\omega_\gamma(\theta=0) \approx 7.0\,\si{\giga\electronvolt}$, as predicted by kinematics (Fig.~\ref{fig:gamma-rays}\,a,\,d)
\begin{equation}
    \hbar\omega_\gamma(\theta) = \frac{4\gamma_0^2\hbar\omega_0'}{1+a_0'^2+2\eta_0'+\gamma_0^2\theta^2},
    \label{eq:compton}
\end{equation}
where the scattering angle is small $\theta\ll 1\,\si{\radian}$.
For linear polarization, one can replace $a_0'^2\rightarrow \frac{1}{2}a_0'^2$ to account for the cycle average of the vector potential.
Note that the $\gamma$-ray energy is suppressed by photon recoil for a collision with a high center-of-mass energy $\eta_0'\gtrsim 1$. By integrating the $\gamma$-ray spectrum, one can show that the photon yield is $N_\gamma\approx 3.6\times10^6$, of which $2.9\times 10^6$ photons are above 1\,\si{\giga\electronvolt}.
In the regime of linear Compton scattering, the photon yield will scale proportionally with the fine structure constant, normalized amplitude squared, and number of cycles contained within the XUV pulse~\cite{heinzl2020}.
For a trailing beam containing $N_e \approx 3\times 10^{10}$ electrons, this corresponds to a yield of $\mathcal{N}=N_\gamma/N_e\approx 0.012\,\%$ per electron, which can be further increased by employing a train of XUV pulses (see Discussion).

The driving and trailing beams, which are assumed to be identical, are required to be wider than the initial laser waist in order to efficiently accelerate the laser pulse. As a consequence, the electrons are well collimated $\sigma_\vartheta=\epsilon_n/\gamma_0\sigma_r\approx 3.7\,\si{\micro\radian}$ at a given value of the transverse normalized emittance $\epsilon_n=1\,$\si{\milli\meter}-\si{\milli\radian}.
In the regime of linear Compton scattering, the divergence of the total $\gamma$-ray beam is controlled by the relativistic emission angle $\vartheta\sim 1/\gamma_0\approx 51\,\si{\micro\radian}$; recoil from photon emission will affect the electron momentum and angular distribution of any subsequent emitted photons.

\begin{figure}[t]
    \includegraphics[width=\linewidth]{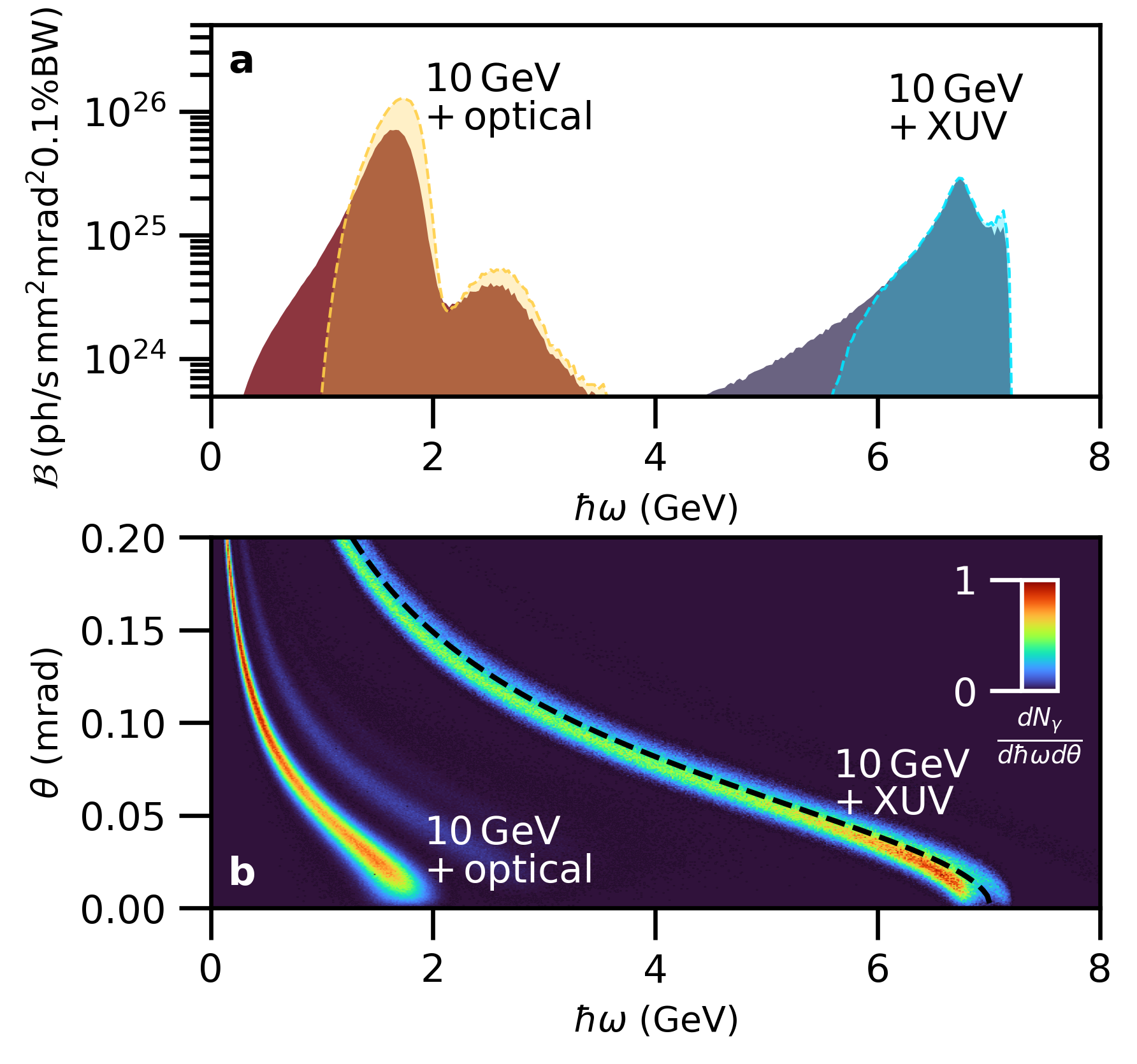}
    \caption{Comparison of collision with optical and XUV laser pulse. (a) Peak brilliance of total $\gamma$-ray beam (dark red and blue) and $\gamma$ rays within $\theta<40\,\si{\micro\radian}$ (light yellow and cyan), where performing angular cuts via a collimator will decrease the bandwidth while maintaining a high peak brilliance. (b) Correlation of polar angle and photon energy, which closely tracks the intrinsic Compton-scattering angle in Eq.~\eqref{eq:compton} (black dashed line). The optical-seeded Compton $\gamma$-ray source features a faint second harmonic, as the normalized amplitude is closer to unity at the point of collision. The key impact of photon acceleration is to increase the $\gamma$-ray energy while keeping the electron energy (10\,\si{\giga\electronvolt}) constant, avoiding the need to extend existing electron accelerators.
    }
    \label{fig:brilliance}
\end{figure}

The $\gamma$-ray polarization is characterized by the Stokes parameters $S_1$, $S_2$, and $S_3$, which indicate vertical and horizontal linear-polarization, diagonal linear-polarization, and circular-polarization, respectively~\cite{blackburn2023}. 
When scattering off the circularly-polarized XUV pulse, high-energy photons at the Compton edge, which are backscattered in the electron rest frame, undergo a change in helicity due to conservation of angular momentum, while low-energy photons inherit their polarization directly from the XUV pulse~\cite{goldman1969}. As a result, the $\gamma$-ray spectrum has a high degree of right-handed circular polarization at the Compton edge $P_{CP}=\frac{1}{2}(1+\bar{S}_3)\approx 95\,\%$ (Fig.~\ref{fig:gamma-rays}\,b). The remaining discrepancy results from higher-order quantum-mechanical corrections.

When scattering off the linearly-polarized XUV pulse, the angular distribution of photons is no longer axially symmetric (Fig.~\ref{fig:gamma-rays}\,f) and instead contains an elongation perpendicular to the electric field characteristic of dipole emission~\cite{blackburn2023}. The degree of linear polarization $P_{LP}=\frac{1}{2}(1 + \bar{S}_1)\approx 77\,\%$ parallel to the electric field peaks at the Compton edge (Fig.~\ref{fig:gamma-rays}\,e), where the mean Stokes parameter $S_1\approx 0.54$ of the photon distribution agrees with the result from linear QED~\cite{goldman1969}
\begin{equation}
    S_1 = \left(1 + \frac{2\eta_0'^2}{1+2\eta_0'}\right)^{-1}, 
\end{equation}
derived in the limit $a_0'\ll 1$. 
From this equation, it immediately follows that the degree of linear polarization is fundamentally limited by photon recoil in contrast to the case of circular polarization. In all simulations, the expected number of positrons generated via the Breit-Wheeler process was far below unity.

\section{Discussion}

To clarify the impact of photon acceleration, the collision is repeated with the initial circularly polarized optical laser pulse in vacuum~(see Appendix~\ref{ap:brilliance}).
Photon acceleration increases the central frequency by an order of magnitude, shifting the Compton edge from 1.5\,\si{\giga\electronvolt} to 7.0\,\si{\giga\electronvolt}, while a monotonic energy-angle correlation ensures that the $\gamma$-ray energy can be selected by performing angular cuts with a collimator (Fig.~\ref{fig:brilliance}).
Achieving the same Compton edge with the optical laser pulse would require a 24\,\si{\giga\electronvolt} electron beam according to Eq.~\eqref{eq:compton}, well beyond the reach of existing accelerators.
The impact of photon acceleration on the $\gamma$-ray energy is therefore equivalent to a 1.4\,\si{\kilo\meter} extension of a 10-\si{\giga\electronvolt} linear accelerator, assuming an average accelerating gradient of 10\,\si{\mega\electronvolt/\meter}.
Remarkably, the increase in photon energy is achieved with only a modest reduction in peak brilliance, $3 \times 10^{25}\,$photons/\si{\second\,\milli\metre^2\,\milli\radian^2}\,0.1\%\,BW at the Compton edge for the collision with the XUV pulse, as only a fraction of the initial laser photons are captured and accelerated by the wake.

Existing user facilities capable of producing multi-\si{\giga\electronvolt} $\gamma$ rays rely either on coherent bremsstrahlung to achieve partial linear polarization (40\,\%) at 9\,\si{\giga\electronvolt}~\cite{adhikari2021}, or on linear Compton scattering off low-intensity ($a_0\ll 1$) ultraviolet laser light to generate linearly or circularly polarized $\gamma$ rays up to 2.4\,\si{\giga\electronvolt}~\cite{muramatsu2022}.
In both cases, the $\gamma$ ray brilliance is orders of magnitude lower than nonlinear Compton-scattering experiments in the multi-\si{\mega\electronvolt} domain, which employ high intensity ($a_0\gg 1$) optical laser light to achieve a high peak brilliance ($10^{23}$ photons/\si{\second\,\milli\metre^2\,\milli\radian^2}\,0.1\%\,BW at 34\,\si{\mega\electronvolt}~\cite{mirzaie2024}), and photon energies of $\sim 1$\,\si{\giga\electronvolt} in the tail of the synchrotron curve~\cite{matheron2024}.
In these experiments the $\gamma$-ray spectrum is broadband, linearly polarized, with a wide angular divergence $\vartheta\sim a_0/\gamma_0$ that is increased by photon recoil in the high $a_0$ regime~\cite{harshemesh2012, blackburn2018, blackburn2020}.
Other theoretical concepts for the generation of brilliant multi-\si{\giga\electronvolt} $\gamma$ rays include collisions of several multi-\si{\peta\watt} laser pulses~\cite{gonoskov2017, magnusson.prl.2019, marklund2023}, two 30\,\si{\giga\electronvolt} lepton beams~\cite{delgaudio2018}, dense electron beams propagating through a metal foil~\cite{benedetti2018}, and Compton scattering in an x-ray free-electron laser oscillator~\cite{hajima2016_prab},
but these do not offer the same degree of control over $\gamma$-ray polarization as the scheme presented here.
The source proposed in this work relies on a 10-\si{\giga\electronvolt} linac and few-\si{\tera\watt} optical laser, which are available at existing facilities, and by operating in the regime of linear Compton scattering we avoid the need to employ spin-polarized electrons to generate circularly polarized $\gamma$ rays.

In summary, photon acceleration provides access to a quasi-linear ($a_0'\lesssim 1$) yet quantum mechanical ($\eta_0'\gtrsim 1$) regime of Compton scattering, unlocking a high peak brilliance and a high degree of circular or linear polarization in the multi-\si{\giga\electronvolt} domain. The key advance underpinning the high brilliance and high $\gamma$-ray energy is the photon accelerator's ability to produce relativistically intense XUV laser light, increasing the center-of-mass energy of the laser-electron collision while maintaining a normalized amplitude close to unity.

A brilliant, circularly polarized, multi-\si{\giga\electronvolt} $\gamma$-ray source would provide an ideal platform to produce low-emittance spin-polarized positrons in a solid target or to observe vacuum birefringence and dichroism in a strong laser field.
A tunable polarization and narrow bandwidth in the multi-\si{\giga\electronvolt} domain could be used to disentangle photo-production mechanisms for charmed baryons and mesons, including $J/\psi$, where higher $\gamma$-ray energies can be accessed by increasing the plasma propagation distance or electron-beam energy.
One aspect of the $\gamma$-ray source presented here which could be improved is the photon yield, for example, by considering an initial optical laser pulse containing tens of cycles.
In this case, the laser pulse would overlap with several plasma periods, producing a train of XUV pulses, which proceed to collide with the trailing beam and generate a series of $\gamma$-ray flashes.

\begin{acknowledgments}
The authors acknowledge support from the Swedish Research Council (Grant No. 2023-06998 and Huvudsekreterarbidrag of M.M.) as well as computational resources provided by Chalmers e-Commons (C3SE) and National Academic Infrastructure for Supercomputing in Sweden (NAISS). This work was supported by National Science Foundation Grant No. 2512014. S.S.B. was supported by U.S. DOE Office of Science Office of HEP under Contract No. DE-AC02-05CH11231. M.J.Q. acknowledges helpful discussions with Carl A. Lindstrøm and Ryan T. Sandberg. The authors acknowledge all contributors to FBPIC and Ptarmigan.
\end{acknowledgments}

\appendix

\section{Photon acceleration simulations}
\label{ap:photon_acc_sims}

All PIC simulations of photon acceleration were carried out with the spectral quasi-3D code FBPIC~\cite{lehe2016}. The 10-\si{\giga\electronvolt} electron driver had a normalized-emittance $\epsilon_r=1\,\si{\milli\meter}$-\si{\milli\radian} and energy spread $\sigma_\gamma=0.1\,\%$. The laser pulse was focused at the peak of the plasma density profile with waist $w_0=7\,\si{\micro\meter}$ and centered where the plasma-density perturbation was zero $\delta n_p = 0$ in the first wavefront behind the driver.

The longitudinal plasma density $n_p\equiv n_p(z)$ profile consisted of a short ramp up to a peak value of $n_{p0}=8\times10^{19}\,\si{\centi\meter^{-3}}$ followed by a tailored down ramp to prevent dephasing~\cite{sandberg2023_prl, sandberg2023_pre}. A moving window of dimensions $[3\sigma_r, 5\lambda_{p0}]$ along $[r, z]$ with a high number of gridpoints $[400, 6144]$ was required to properly resolve XUV wavelengths, where $\lambda_{p0}\approx3.73\,\si{\micro\meter}$. Two azimuthal modes $N_m=2$ and $[2, 4, 3]$ particles-per-cell along $[r, \theta, z]$ were sufficient for convergence. Each PIC simulation required approximately 40 hours (wall time) on a NVIDIA GH200 GPU.

The amplitude of the XUV pulse emitted from the photon accelerator (after propagating through 4\,\si{\milli\meter} of plasma) is multiplied by $\sqrt{R}$ to account for the imperfect reflectivity of the plasma mirror, assuming a typical value of $R=70\,\%$~\cite{gruse2025, shaw2016}. The laser is initialized via the LASY package~\cite{thevenet2024} in a new PIC simulation where the XUV pulse propagates back through the cold, unperturbed plasma profile until it reaches vacuum; this simulation is performed in order to model the divergence of the XUV pulse, which occurs as if it were propagating through vacuum because the plasma is now very underdense compared to an XUV wavelength, as well as to check that the pulse does not drive a significant wake. The final profile of the pulse is then imported directly into Ptarmigan, where the collision with trailing beam takes place.

\section{Plasma mirror}
\label{ap:plasma_mirror}

The radial electric field of the driver at the center of the beam is given by Gauss' law
\begin{equation}
    E_r(r) = \frac{\rho_0 \sigma_r^2}{\varepsilon_0 r}\left(1 - e^{-r^2/2\sigma_r^2}\right),
\end{equation}
which peaks off axis $|E_{r0}|\approx 0.45\rho_0\sigma_r/\varepsilon_0 \approx 9.1\times 10^{12}\,\si{\volt/\meter}$ at $r_0\approx 1.6\sigma_r$, where $\rho_0=-Q/2\pi\sigma_r^2l$ is the beam charge density; this is more than sufficient to directly ionize~\cite{kim2026} a typical valence electron in a plastic target.
The surface plasma is then formed from the valence electrons $N_v=138$ of the monomer (C$_{22}$H$_{10}$N$_2$O$_5$), which has molar mass $M\approx382\,\si{\gram/\mol}$ and density $\rho\approx 1.42\,\si{\gram/\centi\meter^3}$~\cite{nist}. Here $N_A\approx 6.0\times10^{23}\,/\si{\mol}$ is Avogadro's constant. Hence, the plasma mirror is sufficiently overdense $n_{pm}/n_c'\approx 1.8$ to reflect the XUV pulse, where $n_c'= \varepsilon_0 m_e\omega_0'^2/e^2$ is the critical density. 
As the separation of the driver and XUV pulse is in the order of 10\,\si{\femto\second}, the pre-plasma length will be shorter than the central wavelength $\lambda_0'\approx 80\,\si{\nano\meter}$, ensuring efficient specular reflection provided that the initial laser pulse has a high contrast (below $10^{-6}$) a few picoseconds before the arrival of the main pulse. Relativistic focusing and high harmonic generation are negligible here, as the normalized amplitude is below unity upon incidence.

\section{Collision simulations}
\label{ap:collision_sims}

The collision of the XUV pulse and trailing beam is modeled with the Monte Carlo particle-tracking code Ptarmigan~\cite{blackburn2023}, which includes classical dynamics, nonlinear (i.e. multiphoton) QED processes, and takes into account the spatiotemporal structure of both beams.
As the coherent XUV pulse has $a_0' \lesssim 1$, QED processes that take place inside it have formation lengths comparable in size to the XUV wavelength, and it is necessary to use the locally monochromatic approximation (LMA)~\cite{heinzl2020}.
The probability rates and spectra of QED processes thus depend on two parameters: the local, root-mean-square normalized amplitude $\tilde{a}_\text{rms} = e \sqrt{|\langle A^2 \rangle |\,} / (mc)$ and the local energy parameter $\tilde{\eta} = \hbar \kappa^\mu p_\mu / (m^2c^4)$, where $A^\mu = (0, \bm{A})$ and $\kappa^\mu = \tilde{\omega}/c (1, 0, 0, 1)$ are the vector potential and local wavevector of the pulse, and $p^\mu$ is the four-momentum of the electron~\cite{blackburn2023}. In this way, QED events are delocalized at the scale of the XUV pulse wavelength.

The temporal profile of the XUV pulse (along the laser propagation axis) is imported directly into Ptarmigan, which computes the r.m.s potential $\tilde{a}_\text{rms}$ and local frequency $\tilde{\omega}$ as follows.
A dominant frequency scale $\omega'_0$ is extracted from the Fourier-transformed electric field.
Then switching to phase as a dependent variable, $E_x(\phi)$, the potential follows as $A_x(\phi) = \int_{-\infty}^\phi [E_x(\phi') / \omega'_0] \, d \phi'$.
A Hilbert transform of $A_x$ yields the $A_y$ needed to define the complex-valued analytic potential $A_c = A_x + i A_y = |A_c| \exp(i \Phi)$.
The r.m.s. amplitude is given by $\delta |A_c|$, where $\delta = 1/\sqrt{2}$ ($\delta = 1$) for linear (circular) polarization, and the local frequency is $\tilde{\omega} = \omega'_0 (d\Phi / d\phi)$.
The transverse profile of the XUV pulse is assumed to be Gaussian, with waist $w_0' \approx 30\,\si{\micro\meter}$.
This is much larger than the wavelength and so diffraction effects over the length scale of the collision are small.
The trailing beam is identical to the driver except that it has a Gaussian profile $\sigma_z=l/\sqrt{2\pi}$, with the same total charge and peak density.

Ptarmigan then tracks electrons one by one as they propagate through the XUV pulse.
The trajectories of the electrons are defined by their cycle-averaged values, as the fast oscillation at the scale of the XUV period is absorbed into the QED event generator, and satisfy a ponderomotive force equation (see \cite{blackburn2023} for details).
Monte Carlo sampling of the photon emission rates, using either linearly or circularly polarized LMA results as appropriate, is used to generated output particle spectra, e.g. the distribution of $\gamma$-ray energies $dN_\gamma/d\hbar\omega$.
Ptarmigan has been extensively benchmarked against QED theory calculations for plane-wave, focusing, and chirped pulses~\cite{blackburn.njp.2021,blackburn.epjc.2022,blackburn2023}.

\section{Comparison of Compton scattering with optical and XUV pulse}
\label{ap:brilliance}

The peak brilliance can be calculated as a function of the $\gamma$-ray energy
\begin{align}
    \mathcal{B}(\omega) &= \frac{N_{0.1\%}(\omega)}{\pi^2 \sigma_t \sigma_r^2 \sigma_{\theta}^2(\omega)},
    \label{eq:brilliance}
    \\
    N_{0.1\%}(\omega) &= \frac{\hbar\omega}{10^3}\frac{dN_\gamma}{d\hbar\omega},
    \label{eq:photons_in_BW}
\end{align}
where the angular spread of the $\gamma$-ray beam was taken directly from the Ptarmigan simulation data $\sigma_{\theta}(\omega)=\sqrt{\sigma_{\varphi_x}^2(\omega) + \sigma_{\varphi_y}^2(\omega)}$. In practice, the trailing beam acts as a window function while passing through the XUV pulse, effectively controlling the source size and length $\sigma_t=l/c\sqrt{2\pi}\approx 0.4\,\si{\femto\second}$ of the emitted $\gamma$-ray beam.

To clarify the impact of photon acceleration, we repeated our collision simulations with the initial, circularly polarized optical laser pulse. In this case, the laser is focused directly on the Kapton tape (plasma mirror), in the absence of the driving electron beam or plasma, and is assumed to experience the same reflectivity $R=70\,\%$. Immediately after reflection, the optical pulse strikes the trailing beam, where the $\gamma$-ray brilliance is determined by Eqs.~\eqref{eq:brilliance}-\eqref{eq:photons_in_BW}.

\bibliography{bibliography.bib}

\end{document}